\documentclass[twocolumn]{aastex62}

\usepackage{amsmath,amsbsy}

\received{\today}

\shorttitle{Energy Transfer Rate}
\shortauthors{Bandyopadhyay et al.}

\begin{document}
	
\title{Enhanced Energy Transfer Rate in Solar Wind Turbulence Observed near the Sun from Parker Solar Probe}

\author[0000-0002-6962-0959]{Riddhi Bandyopadhyay}
\email{riddhib@udel.edu}
\affiliation{Department of Physics and Astronomy, Bartol Research Institute, University of Delaware, Newark, DE 19716, USA}

\author[0000-0002-5317-988X]{M.~L. Goldstein}
\affiliation{NASA Goddard Space Flight Center, Greenbelt, MD 20771, USA}
\affiliation{University of Maryland Baltimore County, Baltimore, MD 21250, USA}

\author[0000-0002-2229-5618]{B.~A. Maruca}
\affiliation{Department of Physics and Astronomy, Bartol Research Institute, University of Delaware, Newark, DE 19716, USA}
		
\author[0000-0001-7224-6024]{W.~H. Matthaeus}
\affiliation{Department of Physics and Astronomy, Bartol Research Institute, University of Delaware, Newark, DE 19716, USA}

\author[0000-0003-0602-8381]{T.~N. Parashar}
\affiliation{Department of Physics and Astronomy, Bartol Research Institute, University of Delaware, Newark, DE 19716, USA}	

\author[0000-0003-3414-9666]{D. Ruffolo}
\affiliation{Department of Physics, Faculty of Science, Mahidol University, Bangkok 10400, Thailand}

\author[0000-0002-7174-6948]{R. Chhiber}
\affiliation{Department of Physics and Astronomy, Bartol Research Institute, University of Delaware, Newark, DE 19716, USA}
\affiliation{NASA Goddard Space Flight Center, Greenbelt, MD 20771, USA}

\author{A. Usmanov}
\affiliation{Department of Physics and Astronomy, Bartol Research Institute, University of Delaware, Newark, DE 19716, USA}
\affiliation{NASA Goddard Space Flight Center, Greenbelt, MD 20771, USA}

\author[0000-0001-8478-5797]{A. Chasapis}
\affiliation{Laboratory for Atmospheric and Space Physics, University of Colorado Boulder, Boulder, CO 80303, USA}

\author[0000-0001-8358-0482]{R. Qudsi}
\affiliation{Department of Physics and Astronomy, Bartol Research Institute, University of Delaware, Newark, DE 19716, USA}

\author[0000-0002-1989-3596]{Stuart D. Bale}
\affil{Physics Department, University of California, Berkeley, CA 94720-7300, USA}
\affil{Space Sciences Laboratory, University of California, Berkeley, CA 94720-7450, USA}
\affil{The Blackett Laboratory, Imperial College London, London, SW7 2AZ, UK}
\affil{School of Physics and Astronomy, Queen Mary University of London, London E1 4NS, UK}

\author[0000-0002-0675-7907]{J. W. Bonnell}
\affil{Space Sciences Laboratory, University of California, Berkeley, CA 94720-7450, USA}

\author[0000-0002-4401-0943]{Thierry {Dudok de Wit}}
\affil{LPC2E, CNRS and University of Orl\'eans, Orl\'eans, France}

\author[0000-0003-0420-3633]{Keith Goetz}
\affiliation{School of Physics and Astronomy, University of Minnesota, Minneapolis, MN 55455, USA}

\author[0000-0002-6938-0166]{Peter R. Harvey}
\affil{Space Sciences Laboratory, University of California, Berkeley, CA 94720-7450, USA}

\author[0000-0003-3112-4201]{Robert J. MacDowall}
\affil{Solar System Exploration Division, NASA/Goddard Space Flight Center, Greenbelt, MD, 20771}

\author[0000-0003-1191-1558]{David M. Malaspina}
\affil{Laboratory for Atmospheric and Space Physics, University of Colorado, Boulder, CO 80303, USA}

\author[0000-0002-1573-7457]{Marc Pulupa}
\affil{Space Sciences Laboratory, University of California, Berkeley, CA 94720-7450, USA}

\author[0000-0002-7077-930X]{J.C. Kasper}
\affiliation{Climate and Space Sciences and Engineering, University of Michigan, Ann Arbor, MI 48109, USA}
\affiliation{Smithsonian Astrophysical Observatory, Cambridge, MA 02138 USA.}

\author[0000-0001-6095-2490]{K.E. Korreck}
\affiliation{Smithsonian Astrophysical Observatory, Cambridge, MA 02138 USA.}

\author[0000-0002-3520-4041]{A.~W. Case}
\affiliation{Smithsonian Astrophysical Observatory, Cambridge, MA 02138 USA.}

\author[0000-0002-7728-0085]{M. Stevens}
\affiliation{Smithsonian Astrophysical Observatory, Cambridge, MA 02138 USA.}

\author[0000-0002-7287-5098]{P. Whittlesey}
\affiliation{University of California, Berkeley: Berkeley, CA, USA.}

\author{D. Larson}
\affiliation{University of California, Berkeley: Berkeley, CA, USA.}

\author{R. Livi}
\affiliation{University of California, Berkeley: Berkeley, CA, USA.}

\author[0000-0001-6038-1923]{K.G. Klein}
\affiliation{Lunar and Planetary Laboratory, University of Arizona, Tucson, AZ 85719, USA.}

\author{M. Velli}
\affiliation{Department of Earth, Planetary, and Space Sciences, University of California, Los Angeles, CA 90095, USA}

\author{N. Raouafi}
\affiliation{Johns Hopkins University Applied Physics Laboratory, Laurel, MD 20723, USA}

	
	
\begin{abstract}
Direct evidence of an inertial-range turbulent energy cascade has been provided by spacecraft observations in heliospheric plasmas. In the solar wind, the average value of the derived heating rate near 1 au 
is $\sim 10^{3}\, \mathrm{J\,kg^{-1}\,s^{-1}}$, 
an amount sufficient to account for observed departures from adiabatic expansion. Parker Solar Probe (PSP), even during its first solar encounter, offers the first opportunity to compute, in a similar fashion, a fluid-scale energy decay rate, much closer to the solar corona than any prior in-situ observations. Using the Politano-Pouquet third-order law and the von K\'arm\'an decay law, we estimate the fluid-range energy transfer rate in the inner heliosphere, at heliocentric distance $R$ ranging from $54\,R_{\odot}$ (0.25 au) to $36\,R_{\odot}$ (0.17 au). The energy transfer rate obtained near the first perihelion is about 100 times higher than the average value at 1 au. This dramatic increase in the heating rate is unprecedented in previous solar wind observations, including those from Helios, and the values are close to those obtained in the shocked plasma inside the terrestrial magnetosheath.
\end{abstract}

\keywords{magnetohydrodynamics (MHD) --- plasmas --- turbulence --- (Sun:) solar wind}


\section{Introduction} \label{sec:intro}
A fundamental characteristic of a well-developed turbulent system is the transfer of energy from large length scales to small scale structures, i.e., a turbulent energy cascade. The largest structures, known as the ``energy-containing eddies", can be thought of as a reservoir of energy 
that injects energy into a broadband 
scale-to-scale transfer process. 
Assuming homogeneous and stationary fluctuations at these length scales, 
and a similarity law for turbulence evolution,
K\'arm\'an \& Howarth~\citep{Karman1938PRSL} 
develop
a phenomenology to quantify the energy decay rate 
for moderate to high Reynolds-number flows. 
Here, as a first step, 
we employ a von K\'arm\'an law, generalized to magnetohydrodynamics (MHD)\citep{Wan2012JFM,Bandyopadhyay2018PRX,Bandyopadhyay2019JFM}, for evaluating an approximate estimate of the large-scale energy decay rate close to the Sun.

For the next stage of turbulence transfer, at the inertial-range scales, \cite{Kolmogorov1941c} hypothesized a constant transfer of energy across scales,
once again assuming idealized conditions. 
The associated scale-invariant energy flux has been measured in fluid 
and 
as well as 
MHD simulations~\citep{Verma2004PR}. For in-situ 
observations, the Kolomogorov-Yaglom law, generalized to MHD~\citep{Politano1998GRL, Politano1998PRE}, provides a practical way of computing the inertial-range energy cascade rate, directly. The Politano-Pouquet third-order law~\citep{Politano1998GRL, Politano1998PRE}, has been  established as a 
rather 
satisfactory methodology for space plasma observations~\citep{Sorriso-Valvo2007PRL,MacBride2005SW,MacBride2008ApJ,Stawarz2009ApJ}.
The same formalism, as well as several generalizations, 
have been applied to numerical simulations~\citep{Mininni2009PRE}. 

At scales approaching the ion-gyro radius or ion-inertial length,
an MHD description is not valid. Kinetic effects become important near these small scales in a weakly-collisional plasma such as the solar wind,
and the energy cascade process becomes increasingly more complex.
Several channels of energy conversion become available, dispersive plasma 
processes enter the description, 
and dissipation processes become progressively more important. 
A full understanding of the energy dissipation process and pathways at kinetic scales is far from complete at this stage and beyond the scope of 
the present discussion~\citep{Yang2017PoP,Klein2017JPP,Hellinger2018ApJL,Bandyopadhyay2018bApJ,Chen2019NC}. Nevertheless, if the system is 
evolving sufficiently slowly,
one expects that, on average, the total amount of energy transferred from the energy-containing scales, and cascaded though the inertial scales, would account for the total that is eventually 
emerges as dissipation on protons, electrons, and minor ions \citep[see e.g.,][]{Wu2013PRL,Matthaeus2016ApJL}.

A recent study of turbulence 
in the Earth's magnetosheath 
indeed demonstrates 
that the estimate of MHD-scale energy decay rate derived from the von K\'arm\'an law 
at the large scales, and the estimate from the  
third-order law in the inertial scales, agree very well~\citep{Bandyopadhyay2018bApJ}. 
Here, we carry out a similar third-order 
law estimate of inertial range energy transfer and compare with the von K\'arm\'an decay rate using PSP data from the 
first orbit, closer to the Sun than any previous spacecraft mission. 

In this paper, we will address two main questions: 
First, 
what is the average energy decay rate close to the Sun, within 0.3 au ($60\,R_{\odot}$)? Second, 
is there consistency
between the cascade rate 
obtained at the energy-containing scales, evaluated from the von K\'arm\'an law, and that 
estimated from the third-order law at inertial range scales?
    
Regarding the first point, from theoretical expectations and numerical simulations, there is growing support for substantial plasma heating in 
close proximity of the corona, an
effect that 
is possibly 
a result of a 
turbulent cascade~\citep{Matthaeus1999ApJ,Verdini2009ApJ}. 
However, in the absence of any in-situ measurement closer than 0.3 au, estimates remained indirect, until now. 
During its first solar encounter (E1), 
NASA's Parker Solar Probe (PSP)~\citep{Fox2016SSR}, 
sampled the young solar wind in this uncharted territory of the 
heliosphere, ranging from $54\,R_{\odot}$ to $36\,R_{\odot}$. 
We 
report below evidence for substantial heating of the plasma near PSP's first perihelion. 

Our second main question is a natural consequence of the 
nearly incompressible nature of solar wind turbulence \citep{ZankMatt92b}, 
along with the second Kolmogorov hypothesis: 
The presence of 
a scale-invariant incompressive energy flux for some range of length scales 
suggests an inertial range cascade 
that continues
in a conservative fashion
the transfer initiated at the energy containing scales. 
This Kolmogorov conjecture has been tested 
both
directly and indirectly in MHD fluids, and also in weakly-collisional plasmas. Nevertheless, it has not been 
clear how well these conclusions 
hold 
near 
the corona in the 
very young solar wind.
The present 
analysis of PSP data during the first solar encounter demonstrates a good 
correspondence of the von K\'arm\'an and inertial-range cascade rates 
in the inner-heliospheric solar wind.

\section{Data Selection and Method} \label{sec:overview}
\begin{figure*}
	\begin{center} 
		\includegraphics[width=\textwidth]{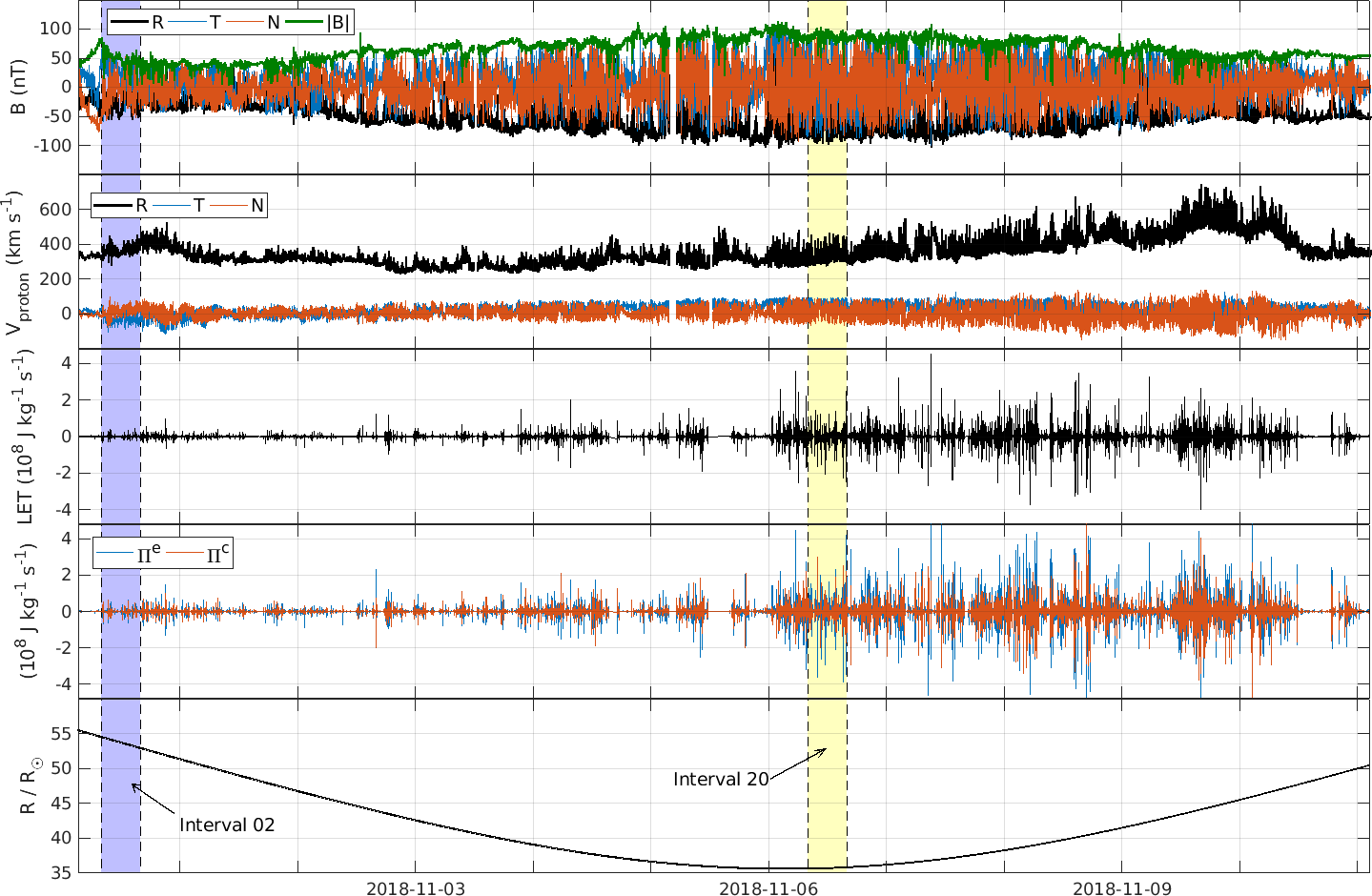}
		\caption{From top: time series plot of magnetic field components, proton velocity components; the local energy transfer (LET) rate proxy, $\Pi (\ell, t)$, obtained from the unaveraged Yaglom law, equation~(\ref{eq:let}), for a lag of $\ell \approx 500\,d_{\mathrm{i}}$; the two terms (equation~(\ref{eq:letec})) of the LET; and the distance of the spacecraft from the Sun, in units of solar radius for the first PSP encounter. The two intervals, discussed in detail in the text, are marked by blue (interval 02) and yellow (interval 20) highlighted parts.}
		\label{fig:overview}
	\end{center}
\end{figure*}
Our analysis 
employs 
magnetic field measurements by the FIELDS~\citep{Bale2016SSR} fluxgate magnetometer (MAG), along with proton velocity and density data from the Solar Probe Cup (SPC)~\citep{Case:SPC}(in this volume) of the SWEAP instrument suite~\citep{Kasper2016SSR}. Although both instruments perform measurements all through the orbits, the data collection rate is relatively low for most of the orbits when the spacecraft is far from the Sun. The primary science-data collection, at a high cadence, occurs during the encounter phase of each orbit at $R \leq 54\,R_{\odot}\, (0.25\,\mathrm{au})$. The first encounter extends from 2018 October 31 to 2018 November 12, with the first perihelion occurring at 03:27 UT on 2018 November 6. The initial and final days do not have full coverage of the high-time-resolution data, so we perform the analysis on data obtained between 2018 November 1 and 2018 November 10. In particular, Level-two (L2) FIELDS and Level-three (L3i) data from the SWEAP archives are used in this paper. The time cadence of the SPC moments varies between 1 NYHz and 4 NYHz, where 1 NYHz (New York Hz) is the inverse of 1 NY sec (New York second), which is approximately equal to 0.874 s~\citep[for an exact definition and more details, see][]{Bale2016SSR}. The native cadence of FIELDS/MAG magnetic field varies from $\approx$ 2.3 to 293 samples per second over E1. To generate a uniform time series, we resample all the variables (SPC and FIELDS) to 1 NYHz cadence. Some spurious spikes in the SPC moments, which are remnants of poor quality of fits, are removed.

To compute 
Elsasser variables from the resampled data, it is necessary to decide
how the density is to be handled to 
produce 
a physically 
meaningful conversion of magnetic field to Alfv\'en speed units. In S.I. units, we have
\begin{eqnarray}
\mathbf{V}_{\mathrm{A}} = \frac{\mathbf{B}}{\sqrt{\mu_{0} m_{\mathrm{p}} n_{\mathrm{p}}}} , \label{eq:valf}    
\end{eqnarray}
where, $\mu_0$ is the magnetic permeability of vacuum, $m_{\mathrm{p}}$ is the proton mass, and $n_{\mathrm{p}}$ is the number 
density of protons. This conversion is to be performed with some care. In strictly incompressible case, $n_{\mathrm{p}}$ is constant, so, in solar wind studies, one typically uses the value of $n_{\mathrm{p}}$ averaged over the whole interval. But, very long-term averages can miss local effects. On the other hand,
single time values may overestimate effects of Alfv\'en speed gradients. Large local variations of density do not imply a possibility of different point-wise Alfv\'en waves. An inertial range
Alfv\'en wave and corresponding Alfv\'en speed should be defined over a reasonably large scale, one 
over which an MHD Alfv\'en wave can exist and propagate. Hence, we use an approach in which the density is averaged over a few correlation times to convert magnetic field fluctuations into Alfv\'enic units.

The correlation time during E1 is around $\tau_{\mathrm{corr}} \sim 300-600\, \mathrm{s}$. These values of correlation time imply that a rolling average of 1250 s covers about 2 to 4 characteristic timescales through E1. Therefore, we use this rolling-averaged density to compute the Alfv\'en velocity, and subsequently, to obtain the Elsasser variables. See figure 1 and related discussions in \cite{Parashar:alfvenicity} for more details.

Another subtle complexity arises due to the presence of alpha particles in the ion population. While solar wind alpha particles are a minority population~\cite[e.g.,][]{Kasper2007ApJ} in terms of their number densities, they are important in terms of their mass contribution~\cite[e.g.,][]{Robbins1970JGR, Kasper2007ApJ, Stansby2019AAP}. So the alpha particles  can have a large effect on the estimation of the denominator in equation~\ref{eq:valf}. Since alpha-particle abundance ratio are currently unavailable, we repeat the analysis assuming a $5 \%$ alpha particle abundance by number. However, repeating the analyses this does not change the results significantly (within $\pm 15 \%$). At this stage, a normalization of the Alfven speed (equation~\ref{eq:valf}) taking into account alpha particle drifts and pressure anisotropies~\citep{Barnes79a, Alterman2018ApJ} is not feasible with the currently available data. However, here we also recall our method of normalizing. Since we are averaging density over roughly a correlation scale, the pressure/temperature anisotropy would also need to be averaged. It is rather certain that the temperature anisotropy will have an average very close to unity when averaged this way, so we anticipate that the effect would be negligible.

We divide E1 $(\sim 10~\mathrm{days})$ into 8 hour subintervals and 
perform 
the analysis on each 
8 hour sample individually, labeling each according to its sequence since the beginning of the encounter (2018 October 31), when the spacecraft first crossed below $54\,R_{\odot}$ (0.25 au). So, for example, interval 2 ranges from 08:00:00 to 16:00:00 UTC, on 2018 October 31 and interval 20 is between 08:00:00 to 16:00:00 UTC, on 2018 November 06. The two subintervals are highlighted in figure~\ref{fig:overview}. These two subintervals are reported and discussed in detail as specific samples for demonstration in the following sections; interval 02 being near the beginning of E1 at $R \approx 54\,R_{\odot}$ and interval 20 near the first perihelion $R \approx 36\,R_{\odot}$. Later, we show the energy decay rates for all the subintervals.

\section{Inertial range} \label{sec:inertial}
To estimate the energy cascade rate at the inertial scales, $\epsilon$, we use 
the Kolmogorov-Yaglom law, extended to isotropic MHD~\citep{Politano1998GRL,Politano1998PRE},
\begin{eqnarray}
Y^{\pm}(\ell) = - \frac{4}{3} \epsilon^{\pm} \ell \label{eq:3ord},
\end{eqnarray}
where, $Y^{\pm}(\ell)=\langle \boldsymbol{\hat{\ell}}\cdot \Delta \mathbf{Z}^{\mp}(\mathbf{r}, \boldsymbol{\ell}) |\Delta \mathbf{Z}^{\pm}(\mathbf{r}, \boldsymbol{\ell})|^2 \rangle$, are the mixed third-order structure functions. Here, $\Delta \mathbf{Z}^{\pm}(\mathbf{r}, \boldsymbol{\ell}) = \mathbf{Z}^{\pm}(\mathbf{r} + \boldsymbol{\ell}) - \mathbf{Z}^{\pm}(\mathbf{r})$ are the increments of Elsasser variables at position $\mathbf{r}$ and 
spatial lag 
$\boldsymbol{\ell}$, and $\langle \cdots \rangle$ denotes spatial averaging. The Elsasser variables are defined as
\begin{eqnarray}
\mathbf{Z}^{\pm} = \mathbf{V} \pm \mathbf{V}_{\mathrm{A}} \label{eq:Z},
\end{eqnarray}
where, ${\bf V}$ is 
the plasma (proton) velocity. The variables, $\epsilon^{\pm}$ in equation~(\ref{eq:3ord}), denote the mean decay rate of the respective Elsasser energies (per units mass): $\epsilon^{\pm} = \mathrm{d} {(Z^{\pm})}^{2} / \mathrm{d}t$; where $Z^{\pm}$ are the root-mean-square fluctuation values of the Elsasser fields. The total (kinetic + magnetic) energy decay rate can be calculated as $\epsilon = (\epsilon^{+} + \epsilon^{-})/2$. For single-spacecraft observations, the structure functions are computed for different temporal lags $\tau$. We then use Taylor's ``frozen-in'' flow hypothesis~\citep{Taylor1938PRSLA} to interpret the temporal lags as spatial lags:\
\begin{eqnarray}
 \boldsymbol{\ell} = - \langle \mathbf{V} \rangle \tau \label{eq:taylor}.
\end{eqnarray}

Here, $\langle \cdots \rangle$ represents temporal averaging, 
further interpreted as spatial averaging using Taylor's hypothesis. The Taylor hypothesis can be applied if the sampled structures convect past the spacecraft sufficiently fast so that the non-linear evolution is negligible during the transition. For MHD-scale turbulent fluctuations, such a condition is met if the mean radial flow speed of the plasma, in the spacecraft frame, is super-Alfv\'enic~\citep{Jokipii1973ARAA}. A comparison of the solar wind speed to Alfv\'en speed for E1 shows that $|\mathbf{V}|/|\mathbf{V}_{\mathrm{A}}| \sim 3-4$, marginally sufficient to use the Taylor hypothesis. A detailed study of Taylor's hypothesis for E1 is reported in \cite{Chasapis:prep} (in this special issue, also see \cite{Parashar:alfvenicity} in this volume).

The presence of a strong mean-magnetic field  in  the  solar  wind creates various kinds of anisotropy~\citep{Horbury2012SSR,Oughton2015PTRSA}. However, in cases where the isotropic PP98 formulation has been compared to a fully anisotropic determination, it is found that the implied cascade rates compare rather well~\citep{Osman2011PRL,Verdini2015ApJ}. Regarding this well-known lack of isotropy in the solar wind, we note that apart from the magnetic field direction, other preferred directions may influence on the turbulence. Notable among these is the radial direction \citep{Volk1973ASS, DongEA14, Verdini2015ApJL} which, due to expansion effects, may violate even the less restrictive assumption of axisymmetry \citep{LeamonEA98-jgr, Chen2012ApJ, Vech2016ApJL, Roberts2017ApJL}. In spite of these caveats, lacking clear alternatives to the use of the isotropic form of the third order relation, we proceed using the Politano-Pouquet formulation as an approximation.

The third-order law (equation~(\ref{eq:3ord})) calculates the \emph{average} energy transfer rate per unit mass, at a given scale $\ell$. Here, the \emph{average} implies 
an ensemble mean,
approximated 
here as an average
along one dimension by the spacecraft.
Recently, the 
local energy transfer rate (LET) dependent 
on scale $\ell$
has been examined
(approximately) 
by using as a surrogate
the \textit{unaveraged} third-order structure function that appears in the third-order law~\citep{Sorriso-Valvo2018SP,Sorriso-Valvo2018JPP,Kuzzay2019PRE},
\begin{eqnarray}
\mathcal{Y}^{\pm}(\ell,\mathbf{r}) &=& - \frac{4}{3} \Pi^{\pm}(\ell,\mathbf{r}) \ell \label{eq:let},
\end{eqnarray}
where, $\mathcal{Y}^{\pm}(\ell,\mathbf{r}) = \boldsymbol{\hat{\ell}}\cdot \Delta \mathbf{Z}^{\mp}(\mathbf{r}, \boldsymbol{\ell}) |\Delta \mathbf{Z}^{\pm}(\mathbf{r}, \boldsymbol{\ell})|^2 $ is the unaveraged kernel of the third-order law (equation~(\ref{eq:3ord})). Now, $\Pi^{\pm}(\ell,\mathbf{r})$ acts as a proxy for the respective Elsasser energy flux, for a lengthscale of magnitude $\ell$, at a position of $\mathbf{r}$. Alternatively, for single-spacecraft observations we can express this as $\Pi^{\pm}(\ell,t)$, where the time $t$ is related to $\mathbf{r}$ in a solar wind frame by Taylor's hypothesis (equation~(\ref{eq:taylor})). As usual, the total rate of local transfer of energy is given by averaging the LET for the two Elsasser fields:
\begin{eqnarray}
\Pi = \frac{1}{2}(\Pi^{+} + \Pi^{-}) \label{eq:letpm}.
\end{eqnarray}
When averaged over a sufficiently large number of samples, for different positions $\mathbf{r}$ (or time $t$), $\Pi$ reduces to the energy flux $\epsilon$. The LET can also be written as
\begin{eqnarray}
\Pi = \frac{1}{2}(\Pi^{\mathrm{e}} + \Pi^{\mathrm{c}}) \label{eq:letec},
\end{eqnarray}
where, $\Pi^{\mathrm{e}} = (-3/4\ell)[2\,\boldsymbol{\hat{\ell}}\cdot \Delta \mathbf{V} (\Delta \mathbf{V}^2 + \Delta \mathbf{V}_{\mathrm{A}}^2 )]$, is associated with the energy advected by the velocity, and $\Pi^{\mathrm{c}} = (-3/4\ell)[-4\,\boldsymbol{\hat{\ell}}\cdot \Delta \mathbf{V}_{\mathrm{A}} (\Delta \mathbf{V} \mathbf{\cdot} \Delta \mathbf{V}_{\mathrm{A}} )]$, is associated with the cross-helicity coupled to the longitudinal magnetic field.

\begin{figure}
	\begin{center}
	    \includegraphics[width=0.9\linewidth]{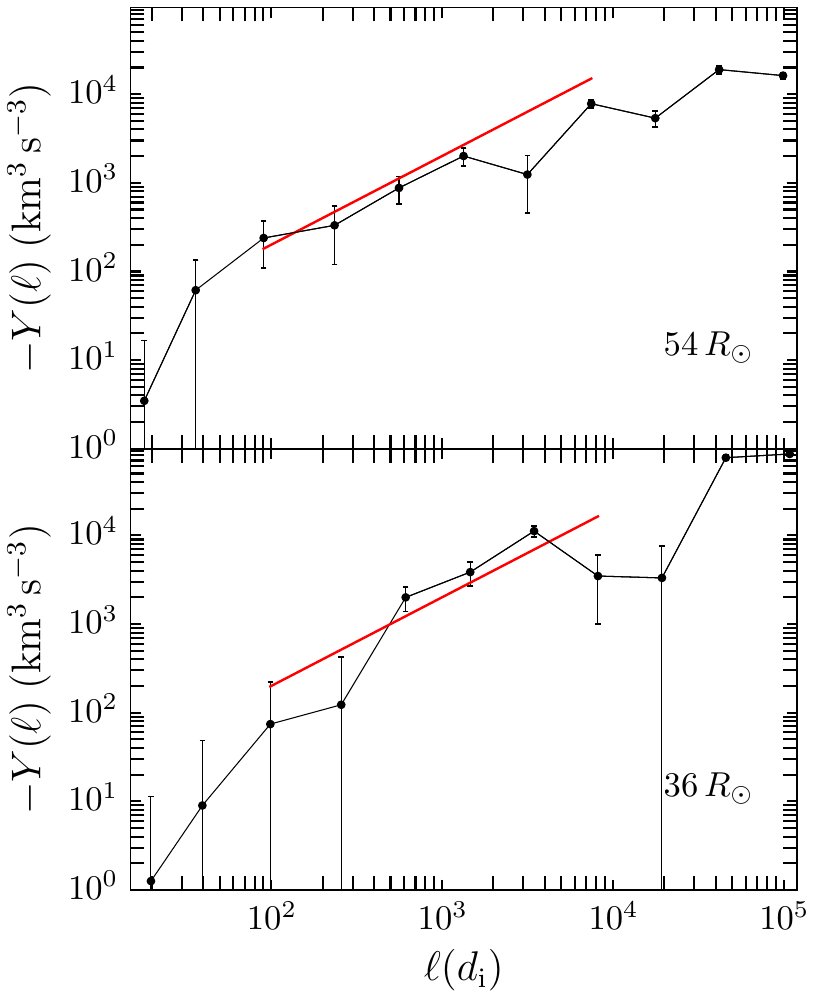}
		\caption{The scaling behavior of the third-order structure function as a function of the (Taylor shifted) length scale in units of ion-inertial length for two different periods we examined. Top: interval 02 on 2018 October 31, from 08:00:00 to 16:00:00 UTC; $R=54~R_{\odot}$, Bottom: interval 20, near the Perihelion, 2018 November 06, from 08:00:00 to 16:00:00 UTC; $R=36~R_{\odot}$. The thick red lines correspond to linear scaling laws for reference. The error bars are calculated by $\sigma/\sqrt{m}$, where $\sigma$ is the standard deviation and $m$ is the number of points used to calculate the mean.}
		\label{fig:ypm}
	\end{center}
\end{figure}

\begin{table}
	\caption{Inertial range cascade rate estimates}
	\label{tab:eps2}
	\begin{center}
		\begin{tabular}{c c c c}
			\hline \hline
			Interval 
			& $R$
			& $ \epsilon $   \\
			& \colhead{$(R_{\odot})$}
			& \colhead{($\rm J\,kg^{-1}\,s^{-1}$)}  \\
			\hline
			02 & 54 & $(5.0\pm0.2)\times10^4$  \\
		20 & 36 & $(2.0\pm0.2)\times10^5$  \\
			\hline
		\end{tabular}
	\end{center}
	\tablecomments{These are based on PP98 MHD adaptation of the Yaglom law.}	
\end{table}

The bottom panel of figure~\ref{fig:overview} shows the LET for E1 by PSP. The LET as shown 
is calculated for an arbitrary lag of $\ell \approx 500\,d_{\mathrm{i}}$, which is well within the inertial range of scales. The proton-inertial length, $d_{\mathrm{i}}$, is defined as $d_{\mathrm{i}} = c / \omega_{\mathrm{p i}} = \sqrt{m_{\mathrm{p}} \epsilon_{0} c^{2} / n_{\mathrm{p}} e^{2}}$, where $c$ is the speed of light in vacuum, $\omega_{\mathrm{p i}}$ is the proton plasma frequency, $\epsilon_{0}$ is the vacuum permittivity, and $e$ is the proton charge. This choice of lengthscale $(\approx 500\,d_{\mathrm{i}})$ 
has no specific 
motivation
other 
than being far from the kinetic $(\sim \, d_{\mathrm{i}})$, as well as the energy-containing scales $(\sim 10^{4}\,d_{\mathrm{i}})$. Use of 
other lags in the inertial range produces
similar results. From the time series of LET, it is clear that the local energy flux is highly intermittent, with sharp variations in fluctuation values, both positive and negative. Note that there is a striking asymmetry in the LET fluctuations, before and after the first perihelion on November 6, 2018. This asymmetry is also observed in other PSP E1 studies and is primarily attributable to 
changes 
in the nature of the solar wind 
during the encounter. 
The solar wind velocity is remarkably slow in the first half and more closely resembles fast solar wind in the second half. We note here that the alpha particle abundances are known to vary with the wind speed~\citep{Kasper2007ApJ}. So the Elsasser variables in the fast wind intervals may be subjected to a larger error due to higher Alpha abundances than in the slow wind. However, as mentioned earlier in the paper, a $5\%$ alpha population does not change the results significantly, so we do not expect that the differences in the inbound and outbound regions are due to underestimating the mass.

If there exists a constant energy flux across the inertial range of length  scales, the mixed, third-order structure function on the left-hand side of equation~(\ref{eq:3ord}) is proportional to the length scale $\ell$ on the right-hand side. Therefore, by fitting a straight line to equation~\ref{eq:3ord}, for a suitable range of scales, one can extract the energy cascade rate. Figure~\ref{fig:ypm} plots the third-order structure function $(Y = (Y^{+} + Y^{-})/2)$ against length scale in units of proton-inertial scale $(d_{\mathrm{i}})$ for two 8 hour subintervals. The top panel shows the structure function for the second subinterval 
in E1 (2018 Oct 27 - 2018 Nov 16), when the spacecraft 
distance ranged from 0.25 au to 
0.17 au from the Sun. A linear scaling is shown for reference. By fitting a straight line, the value of $\epsilon$ is obtained as $5\times10^{4}\,\mathrm{ J\,kg^{-1}\,s^{-1}}$, and the uncertainty from the fitting is $2400\,\mathrm{ J\,kg^{-1}\,s^{-1}}$. We repeat the procedure for a subinterval near the perihelion, a distance of $\sim 36\, R_{\odot}$. Again, a rough linear scaling can be observed, yielding a decay rate of $2\times10^{5}\,\mathrm{ J\,kg^{-1}\,s^{-1}}$, with an uncertainty of $1.5\times10^{4}\,\mathrm{ J\,kg^{-1}\,s^{-1}}$. We perform the same analysis for each 8 hour subinterval for E1. Not every case exhibits a linear scaling, and for those 7 intervals from a total of 34 intervals, we report only the von K\'arm\'an decay estimate of turbulent energy-dissipation rate, as described in the next section.

\section{Energy containing scale} \label{sec:eps1}
\begin{figure}
	\begin{center}
	    \includegraphics[width=\linewidth]{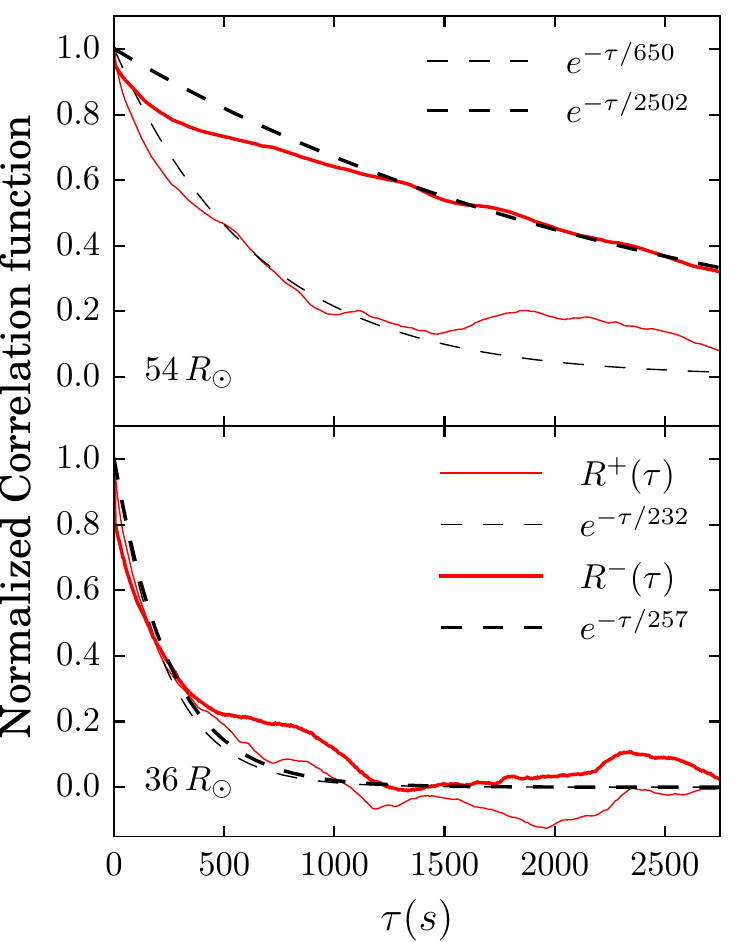}
		\caption{Correlation function versus time lag (seconds) for the Elsasser variables and the corresponding exponential fits for estimation of correlation time. Top: interval 02 on 2018 October 31, from 08:00:00 to 16:00:00 UTC; $R=54\,R_{\odot}$, Bottom: interval 20, near the Perihelion, 2018 November 06, from 08:00:00 to 16:00:00 UTC; $R=36\,R_{\odot}$.}
		\label{fig:corr}
	\end{center}
\end{figure}
We expect that the
global decay rate is controlled, to a reasonable level of approximation, 
by a von K\'arm\'an decay law, generalized to MHD~\citep{Hossain1995PoF,Wan2012JFM,Bandyopadhyay2019JFM}, 
\begin{eqnarray}
\epsilon^{\pm} = -\frac{d (Z^{\pm})^2}{d t} = \alpha_{\pm} \frac{(Z^{\pm})^2 Z^{\mp}}{L_{\pm}}\label{eq:eps1},
\end{eqnarray}
where $\alpha_{\pm}$ are positive constants and $Z^{\pm}$ are the root-mean-square
fluctuation values of the 
Elsasser variables. 

The similarity length scales $L_{\pm}$ in equation~(\ref{eq:eps1}) are the characteristic scales of the 
energy containing eddies. Usually, a natural choice for the similarity scales are 
the associated correlation lengths, 
computed from the two-point correlation functions of the Elsasser variables. Although, there have been some studies showing that this 
association may not always be appropriate~\citep{Krishna-Jagarlamudi2019ApJ}.

The basis of the estimate is determination of the trace of the 
two-point, single-time correlation tensor, 
which 
under suitable conditions,
and assuming the Taylor hypothesis,
is approximately determined by 
the measured two-time correlation. 
The trace of the correlation tensors, computed from 
the Elsasser variables, is given by
\begin{eqnarray}
R^{\pm} (\tau) = \langle \mathbf{Z^{\pm}}(t) \cdot \mathbf{Z^{\pm}}(t+\tau) \rangle_{T} \label{eq:Rzpm},
\end{eqnarray}
where, $\langle \cdots \rangle_{T}$ 
denotes a time average, usually over 
the total time span of the data. 
We use the standard Blackman-Tukey method, 
with subtraction of the local mean, to evaluate equation~(\ref{eq:Rzpm}). 
Although the standard definition of correlation scale 
is given by an integral over the correlation function, 
in practice, especially when there is substantial 
low frequency power present, it is advantageous to 
employ an alternative ``1/e'' definition~\citep{SmithEA01}, 
namely 
\begin{eqnarray}
R^{\pm}(\tau^{\pm}) = \frac{1}{e}\label{eq:corrtau},\\
L_{\pm} = |\langle \mathbf{V}\rangle| \tau^{\pm}\label{eq:corrL},
\end{eqnarray}
where equation~(\ref{eq:corrL}) exploits the Taylor hypothesis. 
Qualitatively, for some spectra with fairly well statistical weight,
the reciprocal correlation length corresponds to the low frequency 
``break" in the inertial range power law. 

\begin{table}
	\caption{Derived variables}
	\label{tab:zpm}
	\begin{center}
		\begin{tabular}{c c c c c c c}
			\hline \hline
			Interval
			& $ Z^{+} $ & $L_{+}$ 
			& $Z^{-}$ & $L_{-}$ & $\sigma_c$ 
			\\
			& \colhead{($\rm km~s^{-1}$)} &  \colhead{($\rm km$)} & \colhead{($\rm km~s^{-1}$)}  & \colhead{($\rm km$)}   
			\\
			\hline
			02 & 88 & $211 \times 10^{3}$ & 54 & $811 \times 10^{3}$ & 0.45 \\
			20 & 126 & $58 \times 10^{3}$ & 48.7 & $65 \times 10^{3}$ & 0.74 \\
			\hline
		\end{tabular}
	\end{center}
	\tablecomments{Elsasser amplitudes $Z^{\pm}$, correlation lengths $L_{\pm}$,
		and normalized cross helicity defined as $\sigma_c = [(Z^{+})^2-(Z^{-})^2]/[(Z^{+})^2+(Z^{-})^2]$.}
\end{table}

Proceeding accordingly, we first compute the Elsasser variables, $\mathbf{Z}^{\pm}$, based on the proton velocity. We then calculate normalized correlation functions for a maximum lag of 1/10\textsuperscript{th} of the total dataset. In Fig.~\ref{fig:corr}, we show the normalized correlation function for each Elsasser variable for the two subintervals at heliocentric distances of $R=54\,R_{\odot}$ and $36\,R_{\odot}$, respectively. Fitting an exponential function to each of the normalized correlation function yields correlation time $\tau^{+} = 585~\mathrm{s}$ and $\tau^{-} = 1786~\mathrm{s}$ for the subinterval at $R=54\,R_{\odot}$, and $\tau^{+} = 585~\mathrm{s}$ and $\tau^{-} = 1786~\mathrm{s}$ for the subinterval at $R=36\,R_{\odot}$. Using the mean flow speed for these two intervals, the deduced correlation lengths are $L_{+} = 211 \times 10^{3} \, \mathrm{km}$ and $L_{-} = 811 \times 10^{3} \, \mathrm{km}$ for interval 02, and $L_{+} = 58 \times 10^{3} \, \mathrm{km}$ and $L_{-} = 65 \times 10^{3} \, \mathrm{km}$ for interval 20. Note that these correlation times are about $10 - 50$ times shorter than the analogous time and length scales for 1~au solar wind~\citep{Matthaeus1982aJGR}, as anticipated by observations and theory \citep[e.g.,][]{Breech2008JGR,Ruiz2014SP,Zank2017ApJ}. We report these lengthscales along with some other obtained quantities in table~\ref{tab:zpm}.

\begin{table}
	\caption{Global decay rate estimates from von K\'arm\'an law}
	\label{tab:eps1}
	\begin{center}
		\begin{tabular}{c c c c}
			\hline \hline
			Interval 
			& $R$
			& $ \epsilon $   \\
			& \colhead{$(R_{\odot})$}
			& \colhead{($\rm J\,kg^{-1}\,s^{-1}$)}  \\
			\hline
			02 & 54 & $(3.0\pm0.7)\times10^4$  \\
		20 & 36 & $(2.00\pm0.07)\times10^5$  \\
			\hline
		\end{tabular}
	\end{center}
\end{table}

Using the obtained Elsasser amplitudes and correlation lengths, we calculate the estimates of of energy decay rate from the von K\'arm\'an law, as outlined in equation~(\ref{eq:eps1}). We use $\alpha_{+} \approx \alpha_{-} = 0.03$. These values are derived from the dimensionless dissipation rate, $C_{\epsilon}$, in MHD turbulence: $\alpha_{\pm} = 2 C_{\epsilon}/ 9\sqrt{3}$~\citep{Usmanov2014ApJ}. The precise values of $\alpha_{\pm}$ depend on several parameters, e.g., the mean-magnetic-field strength, cross helicity, magnetic helicity, but usually remain close to the generic values used here~\citep{Matthaeus2004GRL,Linkmann2015PRL, Linkmann2017PRE,Bandyopadhyay2018PRX}.

The largest source of uncertainty in these calculations is due to the assumption of homogeneity. To estimate the uncertainties associated with each of the von K\'arm\'an decay estimates, we divide each 8 hour interval further into two 4-hour parts and calculate the amplitude of the Elsasser variables in the two halves. From those values of Elsasser amplitude, we calculate the average energy decay rate, and the difference with the original decay rate, estimated from the 8 hour interval, is reported as the uncertainty. The values of the energy decay rate $\epsilon$ 
found in this analysis, for the two subintervals,
are reported in table~\ref{tab:eps1}. A comparison with table~\ref{tab:eps2} affirms that the estimated values are fairly close.


\section{Estimates from Global Simulation} \label{sec:sim}
\begin{figure}[ht!]
	\begin{center}
		\includegraphics[width=\linewidth]{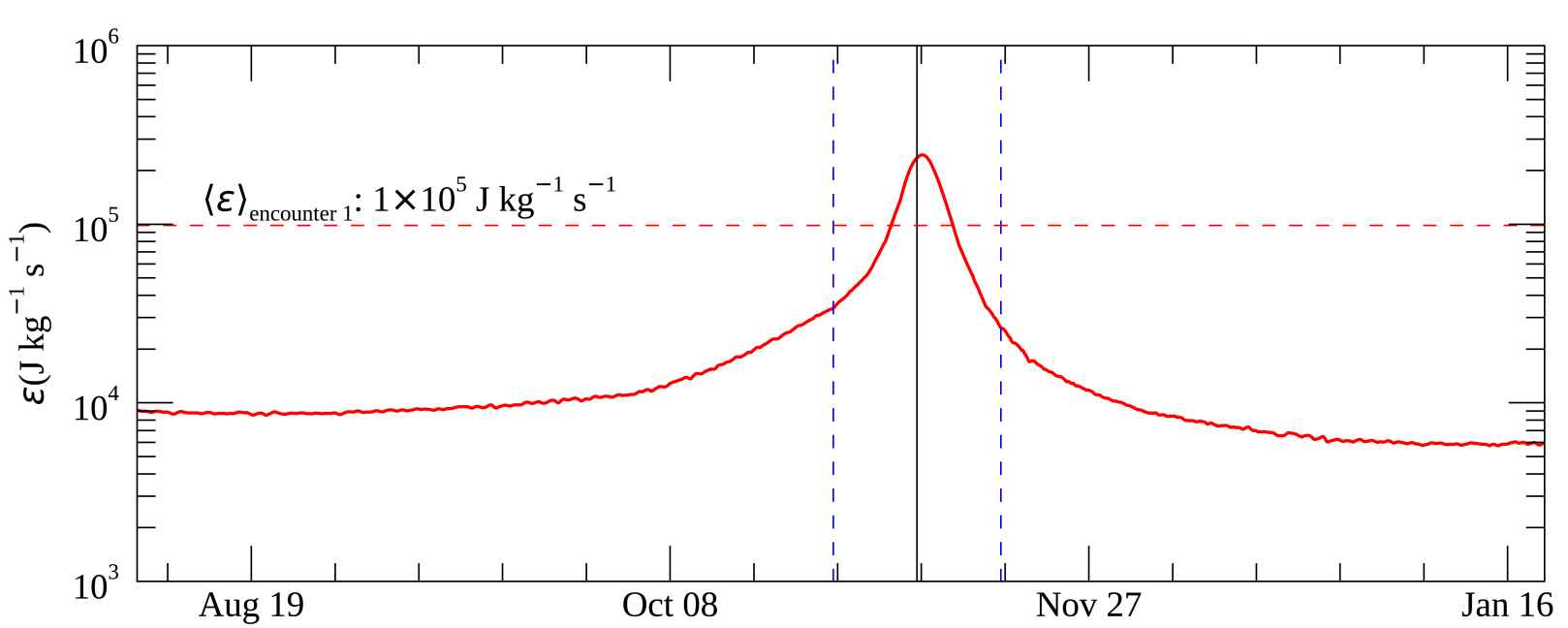}
		\includegraphics[width=\linewidth]{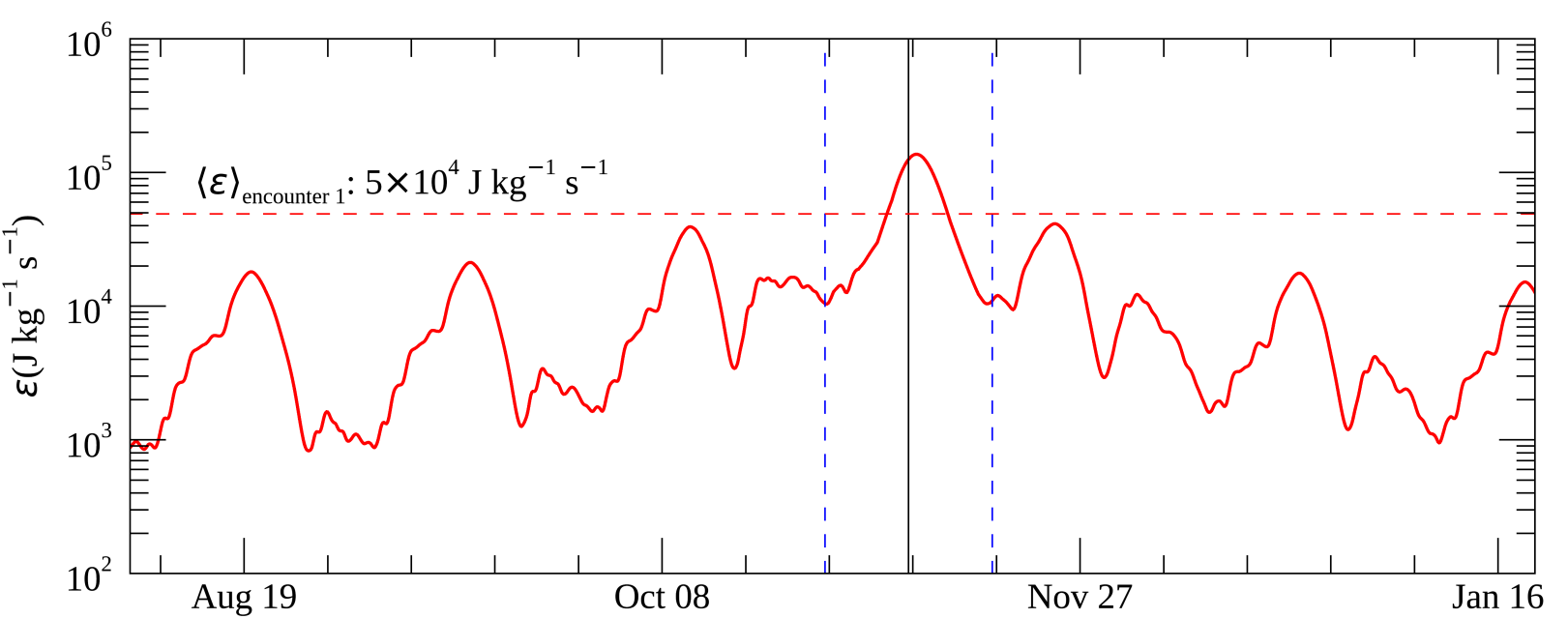}	
		\caption{Results from a global simulation. Top: Untilted dipole simulation, bottom: 2018, November magnetogram simulation. The first solar encounter (E1) is highlighted as the region within blue dashed lines and the first perihelion is shown as the solid vertical line. The average heating rate during the encounter is indicated by a dashed, horizontal line in each case.}
		\label{fig:sim}
	\end{center}
\end{figure}
In this section, we compare the PSP measured cascade rates, as derived in the 
earlier sections, with global solar corona and solar wind  simulations~\citep{Usmanov2018ApJ,Chhiber2019bApJS}. The simulations are three-dimensional, MHD-based models with self-consistent turbulence transport and heating. For comparison with PSP observation, we employ two simulation runs, distinguished by the magnetic field boundary condition at the coronal base. In the first 
run, a Sun-centered, untilted dipole magnetic field is used for the inner surface magnetic field boundary condition. A zero or small tilt of dipole magnetic field with respect to the solar rotation axis is often associated with the conditions of solar minimum, which PSP sampled during its first encounter. The second run is obtained by using the magnetic field boundary condition from 2018 November magnetogram data (Carrington Rotation 2210), normalized appropriately. More details about these simulations can be found in \cite{Chhiber2019bApJS}.

Figure~\ref{fig:sim} shows a profile of heating rates computed using a K\'arm\'an-Howarth like phenomenology from the 
global heliospheric simulations. Here, the simulated data are sampled along a trajectory similar to that of PSP
using an untilted dipole (top panel), and 
using a November 2018 magnetogram (bottom panel). 
The average values for the encounter  
are shown as horizontal lines. 
It is apparent that the expectations based
on the simulations correspond well to the observed values. E1 is indicated by the two dashed vertical lines and the solid vertical line marks the first perihelion.

\begin{figure*}
	\begin{center}
		\includegraphics[width=\linewidth]{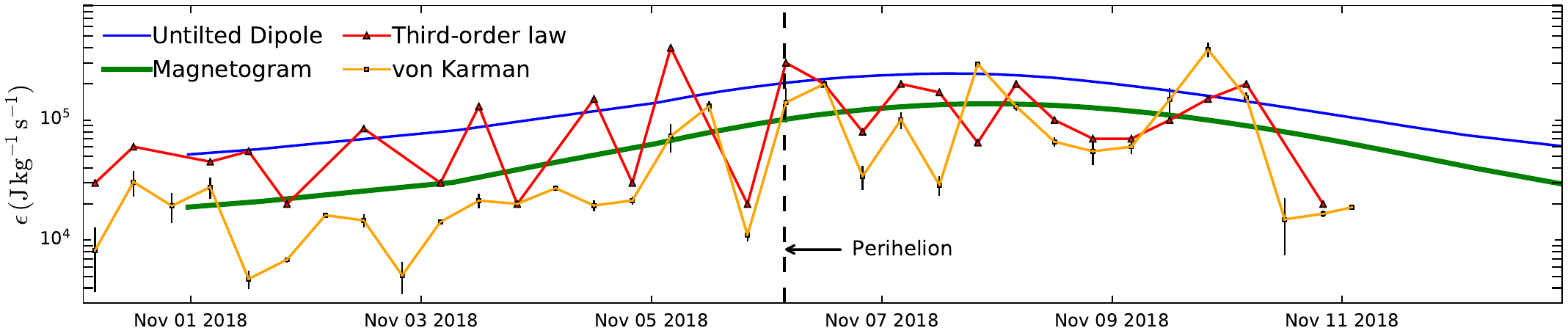}
		\caption{Estimates of the energy decay rate from third-order law (red line with triangles) and  von K\'arm\'an decay rate (orange line with black squares) for E1. The error bars, when they are larger than the symbols, are shown as black, vertical bars. Also shown are the heating rate estimates from a global 
        simulation model along a virtual PSP trajectory, for an untilted dipole condition (thin blue line) and a 2018 November magnetogram simulation (thick green line). The perihelion is shown as a dashed vertical line.}
		\label{fig:e1}
	\end{center}
\end{figure*}

For a more direct comparison of the two kinds of estimates from PSP results, along with the simulation predictions, in figure~\ref{fig:e1}, we plot the different estimates of energy decay rate for E1. The first perihelion is shown as a dashed vertical line, for reference. The thin, blue line and the thick, green line represent the simulation results for boundary condition of a dipole magnetic field with $0^{\circ}$ tilt with respect to the solar rotation axis and that obtained from a 2018 November magnetogram data. The red line with triangular symbols are the values obtained from the third-order law, whenever a linear scaling is found. The orange line with square symbols are calculated using von K\'arm\'an phenomenology.

Evidently, there is a fairly good level of agreement among the different estimates of heating rate. The average heating rate is maximum near the perihelion but is generally higher for the second half of the encounter and the consistency is also better during this period. Possible explanations for this observation are discussed later.

\section{Conclusions} \label{sec:conc}
The first orbit by PSP, particularly when the spacecraft was closer than 0.25~au to the Sun, presents several novel observations, e.g., co-rotational flows~\citep{Kasper2019a:prep}, rapid polarity of reversals in a mostly radial magnetic field (the so-called ``switchbacks"~\citep{Bale:prep}, (Dudok de Wit et al. this volume), numerous sharp, discrete Alfv\'enic impulses  with an anti-Sunward propagating direction (the so-called ``spikes"~\citep{Horbury:prep} this volume), suprathermal 
population~\citep{McComas2019Nature}. In this paper, taking advantage of this unique opportunity to make
{\it in-situ} observations
at unprecedented 
close distances to the 
solar corona,
we have performed some first basic turbulence cascade
rate statistics, analyzing the full E1 data as an ensemble of fluctuations. 

Although much emphasis
is often placed on the 
significance of the
spectral slope ($E(k) \sim k^{-5/3}, k^{-3/2}$), 
the 
scaling of mixed, third-order structure functions~(equation~(\ref{eq:3ord})) is considered a more direct indication of inertial-range cascade in turbulent plasmas. For most of the intervals analyzed here, the averaged third-order structure functions exhibit linear scaling with length scale.

We use two estimates of energy decay rate at the MHD scales: a von K\'arm\'an decay law and a Politano-Pouquet third-order law. The two 
estimates 
are fairly consistent. However, from figure~\ref{fig:e1}, clearly, the agreement is better for the outbound leg of E1. This effect may be attributed to the fact that, on overall the fluctuation amplitudes are higher in the fast solar wind plasma for the later part of E1. The local energy transfer rate is larger in the parts which appear to be more Alfv\'enic. The region after 2018 Nov 09 has a reasonably low speed (under $400\,\mathrm{km\,s^{-1}}$ for the  first part) but high Alfv\'enicity, indicating that some part of this is a sample of Alfv\'enic slow wind~\citep{DAmicis2019MNRAS}. Typically, the slow solar wind is observed to be  more intermittent (e.g. Bruno et al. 2003). The LET has been shown to be related to the partial variance of increments PVI statistics~\citep{Sorriso-Valvo2018SP}. If one assumes that the slow wind is populated with more coherent structures, then the larger fluctuations of LET seen in the fast-wind periods are most likely due to Alfv\'enic fluctuations. We note however that the Alfv\'enicity is not perfect, and that turbulence may be fairly active in these regions since the required mixture of + and - Elsasser amplitudes is present in these epochs~\citep{Parashar:alfvenicity}. Additional studies are required to reach a definite conclusion.

We also see that the \cite{Usmanov2018ApJ} global-simulation model predicts the heating rate reasonably well in comparison
with the PSP observations. Some contextual comparisons with results from previous in-situ observations are also in order. We have mentioned previously that the average cascade rate for 1 au solar wind plasma is approximately $\sim 10^{3}\, \mathrm{J\,kg^{-1}\,s^{-1}}$, the particular value depending on the specific solar wind conditions~\citep{Coburn2014PRS}. Using Helios 2 data, \cite{Hellinger2013JGR} evaluates the heating rate at 0.3 au slow solar wind as $\sim 10^{-15}\,\mathrm{J\,m^{-3}\,s^{-1}} \equiv 6000\, \mathrm{J\,kg^{-1}\,s^{-1}}$, which is close to the value of $\sim10^{4}\, \mathrm{J\,kg^{-1}\,s^{-1}}$ obtained at 0.25 au in the slow wind by PSP, near the beginning of E1. The energy decay rate becomes progressively larger closer to the perihelion and the highest obtained values are almost comparable to the values obtained for the shocked plasma in Earth's magnetosheath~\citep{Bandyopadhyay2018bApJ,Hadid2018PRL}. This
large value of energy cascade rate is presumably due to the strong driving process occurring closer to the corona. There are some peculiarities about E1. These studies, unlike the ones with Helios or Wind data, do not have the statistical capability to investigate underlying parametric dependences. Incorporating more data from future PSP orbits and Solar Orbiter~\citep{Muller2013SP}, capturing more kinds of solar wind, will make the scenario clearer. 

The theory of MHD turbulence, and consequently, the phenomenologies discussed here, are statistical in nature~\citep{MoninYaglom-vol1,MoninYaglom-vol2}. Accordingly, the heating rates obtained here are independent of 
any specific dissipation mechanism and are not applicable to individual events. However, a model of stochastic heating mechanism~\citep{Martinovic:prep} in this volume, accounts for a significant fraction of the heating rates reported in this study, 
suggesting that
stochastic heating may be 
a major contribution to 
overall proton heating in these regions. 
Future 
comparisons with other processes, e.g., reconnection, ion-cyclotron damping, Landau damping, will be useful, using measurements of the particle Velocity Distribution Function (VDF)~\citep{He2015ApJL} or methods such as Field-Particle correlations~\citep{Klein2016ApJL}.

The present study is not without limitations and those deserve to be addressed at this point. The two decay laws we employed, 
are derived on the basis of homogeneity and stationarity~\citep{Politano1998GRL,Politano1998PRE,Wan2012JFM}. Whether solar wind fluctuations, particularly the ones 
sampled by PSP in the inner heliosphere, satisfy the conditions of weak stationarity, is not entirely clear and requires more rigorous investigation~\citep{Matthaeus1982bJGR} (\cite{Parashar:alfvenicity}, \cite{Chhiber:pvi_WT}, Dudok de Wit et al., this volume). In regions close to the corona, expansion effects may become important, but they are ignored in the two phenomenologies used here. Finally, 
we have not explored the statistics of LET here. More detailed studies that pursue these considerations await.

\section*{Acknowledgments}
 Parker Solar Probe was designed, built, and is now operated by the Johns Hopkins Applied Physics Laboratory as part of NASA’s Living with a Star (LWS) program (contract NNN06AA01C). Support from the LWS management and technical team has played a critical role in the success of the Parker Solar Probe mission.
This research was partially supported by the Parker Solar Probe Plus project through Princeton/IS$\odot$IS subcontract SUB0000165, NASA grant 80NSSC18K1210, NSF-SHINE AGS-1460130, and in part by grant RTA6280002 from Thailand Science Research and Innovation. S.D.B. acknowledges the support of the Leverhulme Trust Visiting Professorship program.


\end{document}